\documentclass[a4paper]{article}

\usepackage{INTERSPEECH2021}

\title{WNARS: WFST based Non-autoregressive Streaming End-to-End Speech Recognition }
\name{Zhichao Wang, Wenwen Yang, Pan Zhou, Wei Chen}
\address{
  AI Interaction Division, Sogou Inc., Beijing, P.R.China 
}
\email{\{wangzhichao214232,yangwenwen,zhoupan,chenweibj8871\}@sogou-inc.com}

\begin{document}
\ninept
\maketitle
\begin{abstract}
  Recently, attention-based encoder-decoder (AED) end-to-end (E2E) models have drawn more and more attention in the field of automatic speech recognition (ASR). AED models, however, still have drawbacks when deploying in commercial applications. Autoregressive beam search decoding makes it inefficient for high-concurrency applications. It is also inconvenient to integrate external word-level language models. The most important thing is that AED models are difficult for streaming recognition due to global attention mechanism. In this paper, we propose a novel framework, namely WNARS, using hybrid CTC-attention AED models and weighted finite-state transducers (WFST) to solve these problems together. We switch from autoregressive beam search to CTC branch decoding, which performs first-pass decoding with WFST in chunk-wise streaming way. The decoder branch then performs second-pass rescoring on the generated hypotheses non-autoregressively. On the AISHELL-1 task, our WNARS achieves a character error rate of 5.22\% with 640ms latency, to the best of our knowledge, which is the state-of-the-art performance for online ASR. Further experiments on our 10,000-hour Mandarin task show the proposed method achieves more than 20\% improvements with 50\% latency compared to a strong TDNN-BLSTM lattice-free MMI baseline.
  
\end{abstract}

\section{Introduction}
\label{sec:intro}
Traditional automatic speech recognition (ASR) systems usually consist of three components: acoustic model (AM), pronunciation model (PM) and language model (LM). End-to-end (E2E) ASR systems provide an elegant way to combine these components into a single network and significantly simplify the training process.  There are three major types of E2E architectures in ASR: connectionist temporal classification (CTC) \cite{graves2006connectionist, amodei2016deep, miao2015eesen}, recurrent neural network transducer (RNN-T) \cite{graves2013speech, he2019streaming, rao2017exploring} and attention based encoder-decoder (AED) \cite{chorowski2014end, chan2015listen, chorowski2015attention}. Recent study proposed a hybrid CTC-attention architecture which combines CTC and AED into a single model and has achieved promising results in many ASR tasks \cite{kim2017joint,hori2017advances}.

The CTC-attention architecture introduces a CTC objective to the encoder during training of AED models, which encourages the alignments between the inputs and outputs of AED models to be monotonic \cite{kim2017joint}. During inference, a one-pass decoding method is proposed, which combines CTC, attention and an additional recurrent neural network language model (RNN-LM) to generate final results \cite{hori2017advances}. While this hybrid architecture has made significant progress in ASR, there are three challenges to tackle in practical applications:

\begin{enumerate}
\item Decoding efficiency: The CTC-attention model is decoded in an autoregressive way. Each output token is generated iteratively by conditioning on previously generated tokens. The decoder has to compute as many times as the output length, which is time consuming.
\item LM integration: Also, the CTC-attention model is inconvenient to take advantage of word-level LMs due to the autoregressive problem. In \cite{hori2017advances}, a character-level RNN-LM is integrated to do one pass decoding. However, in speech recognition, word-level LMs are much more powerful than character-level LMs.
\item Streaming decoding: Since the input and output sequences of AED models are of different lengths, the attention module inside decoder has to pay attention to the entire encoder outputs to generate soft alignments between input and output sequences. This makes it difficult to do streaming decoding.
\end{enumerate}

Previous work usually focused on solving one of these problems. For example, in \cite{higuchi2020mask}, a two-pass decoding framework is proposed to achieve non-autoregressive decoding. The target sequence is initialized with the greedy CTC outputs and low-confidence tokens are masked based on the CTC probabilities. These masked tokens are then predicted by decoder conditioned on the entire input sequence in a few fixed steps. 
In \cite{toshniwal2018comparison, shan2019component, michel2020early}, multiple LM integration schemes have been proposed for E2E models. All these works focus on the character-level LM integration and it is still not clear how to integrate additional word-level LMs into AED models. In \cite{chiu2017monotonic, fan2018online, kim2019attention, zhangstreaming, tsunoo2020streaming}, efforts have been made to convert full sequence soft attention inside the decoder into local chunk-wise attention, which is suitable for streaming decoding. 
The key element of previous streaming AED methods is predicting boundary in encoder output for local attention of each token\cite{chiu2017monotonic, fan2018online, kim2019attention,tsunoo2020streaming} or predicting number of tokens in each fix-length chunk \cite{zhangstreaming} for sub-sentence level attention.

In this work, we adopts the CTC-attention architecture and propose a novel WFST-based non-autoregressive streaming speech recognition (WNARS) framework to solve the above three problems together.
The CTC branch equipped with WFST for better word level language model integration, performs as a first-pass decoder, which is responsible for streaming decoding. Thus we convert a streaming local attention problem to a streaming CTC problem. 
The decoder branch then rescores the generated hypotheses conditioned on the entire encoder outputs in non-autoregressive way.
Experimental results on the AISHELL-1 and our 10,000-hour industrial Mandarin tasks indicate the effectiveness of our method.

\section{Model architecture}

The following subsections first give a brief review of the attention-based encoder-decoder architecture. Then the hybrid CTC-attention training loss is explained.
\subsection{Attention-based encoder-decoder}
\begin{figure}[t]
  \centering
  \includegraphics[width=0.9\linewidth]{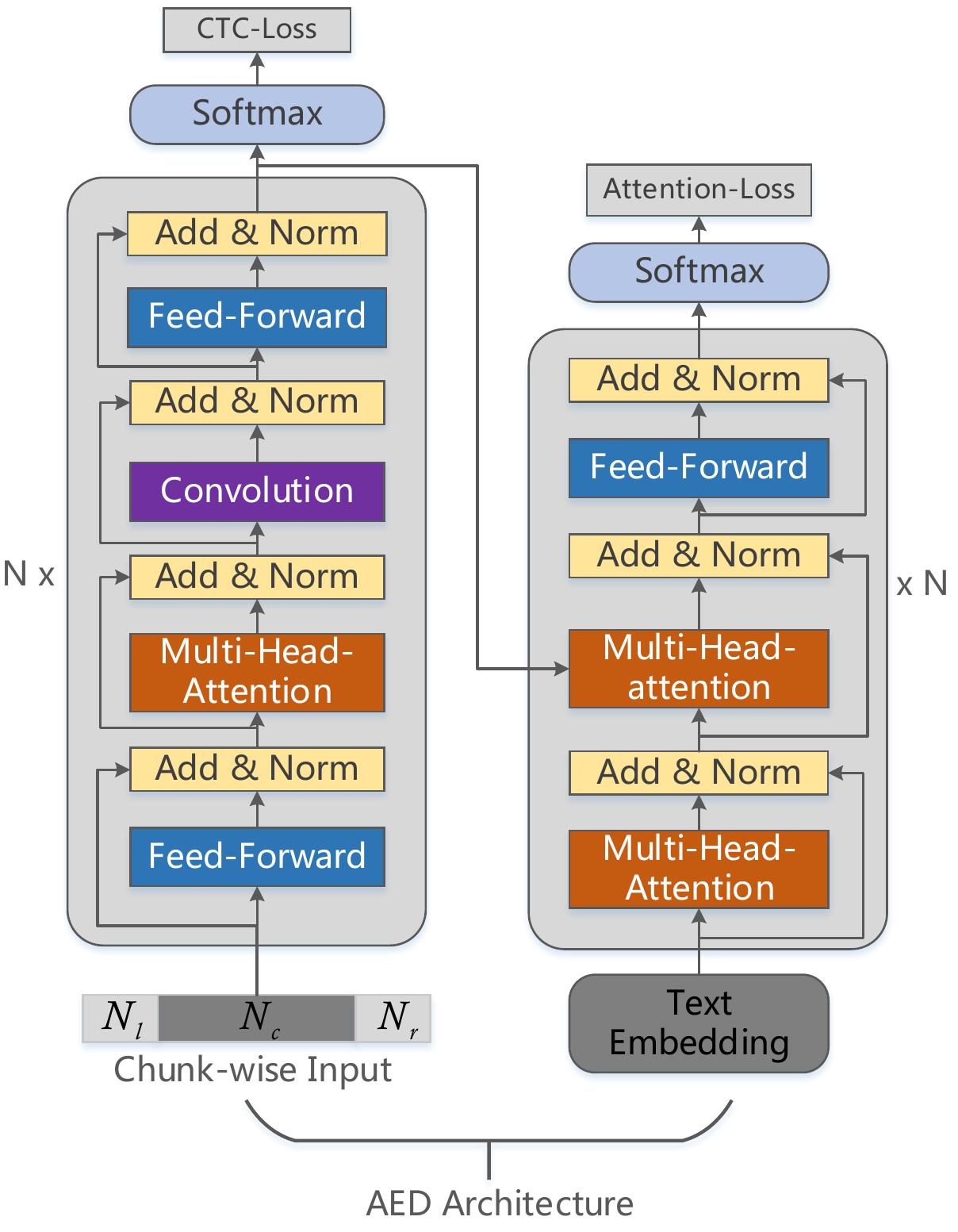}
  \caption{CTC-attention based AED model architecture.}
  \label{fig:conformer}
\end{figure}

The AED architecture consists of an encoder and a decoder. The encoder learns a hidden representation \(\boldsymbol{H}=(h_{u}|u=1,...,U)\) from a speech feature sequence \(\boldsymbol{X}=(x_{t}|t=1,...,T)\), while the decoder generates an output sequence  \(\boldsymbol{Y}=(y_{l}|l=1,...,L)\) based on \boldsymbol{$H$} and previously emitted tokens. It could be formulated as: 
\begin{equation}
  P_{att}(\boldsymbol{Y}|\boldsymbol{X})=\prod_{l=1}^{L}p(y_{l}|y_{1},...,y_{l-1},\boldsymbol{H})
  \label{eq1}
\end{equation}
There are many variants of the AED architecture \cite{chan2015listen, vaswani2017attention, gulati2020conformer}. Our work adopts the Conformer structure as shown in figure~\ref{fig:conformer}. A Conformer encoder-block is composed of four modules, namely, a feed-forward module, a multi-head self-attention module, a convolution module and a second feed-forward module in the end. The multi-head self-attention module captures the global context while the convolution module exploits the local correlations synchronously. A Conformer decoder-block consists of three modules, i.\,e., a feed-forward module, a multi-head self-attention module and a source-target attention module. The source-target attention module captures the alignments between input and output sequences. The encoder-block and decoder-block are stacked multiple times to achieve good performance. 

\subsection{Hybrid CTC-attention loss}

In \cite{kim2017joint}, researchers proposed a hybrid CTC-attention loss to improve the AED model. As shown in figure~\ref{fig:conformer}, a CTC objective function is attached to the shared encoder as an auxiliary task. 
The forward-backward algorithm of CTC can enforce monotonic alignment between speech and label sequences, which helps to reduce irregular alignments of AED models and leads to better performance. The hybrid training loss can be defined as follows:
\begin{equation}
  L_{CTC-attention}=\lambda L_{CTC}+(1-\lambda )L_{attention}
  \label{eq2}
\end{equation}
where $\lambda$ is a hyper-parameter: $0\leq \lambda \leq 1$.

\section{The proposed methods}
In this section, we describe the proposed WNARS framework in detail.

\subsection{WFST-based non-autoregressive decoding}

\begin{figure*}[t]
  \centering
  \includegraphics[width=0.9\linewidth]{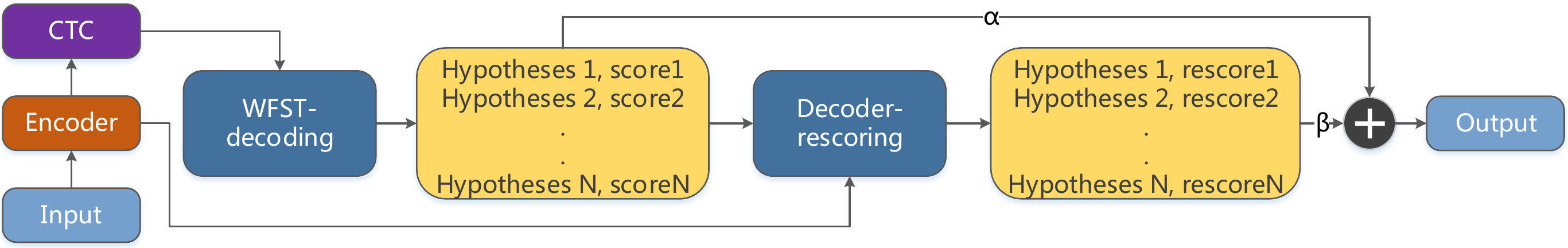}
  \caption{WFST-based non-autoregressive decoding framework.}
  \label{fig:two-pass}
\end{figure*}

 The key to accelerate the decoding efficiency of AED models is to release the model from autoregressive decoding. Fortunately, the CTC-attention architecture gives us a chance to implement non-autoregressive decoding flexibly with a two-pass decoding framework. Like \cite{higuchi2020mask}, we employ CTC to generate initial hypotheses and use attention-decoder to correct the results. Unlike \cite{higuchi2020mask}, during CTC decoding at the first-pass stage, a n-gram word-level LM is integrated by a WFST to generate n-best hypotheses. Then, hypotheses are rescored by the attention-decoder in a teaching-forcing manner at the second pass stage. The two-pass hypotheses scores are combined using tunable hyper-parameters \(\alpha \) and \(\beta \) to generate final results. Figure~\ref{fig:two-pass}. shows the proposed WFST-based non-autoregressive decoding framework. With this framework, we not only bypass the autoregressive problem, but also explore a way to integrate a word-level language model into AED models.

The search graph we used in the first-pass decoding is built by three individual WFSTs as in \cite{miao2015eesen}. A grammar WFST denoted as G encodes the permissible word sequences in a language (word-level language model). A lexicon WFST denoted as L encodes the mapping from sequences of lexicon units (Chinese character in this work) to words. A token WFST denoted as T maps a sequence of frame-level CTC labels to a single lexicon unit. After all, the search graph S is compiled as follows:
\begin{equation}
  S=T\circ min(det(L\circ G))
  \label{eq3}
\end{equation}
where \(\circ\), \(det\) and \(min\) denote composition, determinization and minimization respectively.

\subsection{Chunk-wise streaming processing}
To tackle the streaming problem of AED models, the most common approach is introducing chunk-wise processing for both encoder and decoder. 
However, chunk-wise processing for decoder has to predict boundary in encoder output for local attention of each token or predict number of tokens in each chunk for sub-sentence level attention, which may introduce additional errors.
With the proposed decoding framework described in section 3.1, we only need to do chunk-wise processing for encoder during CTC decoding, leaving the decoder to process global attention during rescoring.
In our framework, the encoder is processed as in \cite{miao2020transformer}, the input sequence is split into isolated chunks of length \(N_{c}\). To compensate contextual information, historical context \(N_{l}\) and future context  \(N_{r}\) are appended to the left and right of  \(N_{c}\), as shown in Figure~\ref{fig:conformer}. The latency of the model is then controlled by  \(N_{c}+N_{r}\). Similar to \cite{zhang2020unified}, We use multi-chunk strategy, which means {\(N_{l}\), \(N_{c}\), \(N_{r}\)} are variable during training. In this way, we can obtain a streaming speech recognizer to satisfy different latency demands with by controlling \(N_{c}\) and \(N_{r}\) during inference.

\section{Experiments}
\subsection{Datasets}
\label{ssec:data}
To evaluate the effectiveness of our proposed methods, we carried out experiments on two Mandarin datasets, namely the open source AISHELL-1 corpus \cite{bu2017aishell} and our own 10,000-hour Mandarin corpus. AISHELL-1 consists of three parts, a 150-hour training-set, a 10-hour development-set and a 5-hour test-set. 
The 10,000-hour corpus consists of 12 million utterances, collected from multi-domain, including dictation, voice-search, natural conversation, education, report et al. The data is anonymized without personal identifiable information. In the 10,000-hour corpus, 100 hours were randomly selected as the development set, a 25-hour close-talk test-set and a 10-hour far-field test-set were used for evaluation.

\subsection{Experimental setup}
We used the CTC-attention based Conformer model in all setups. The input acoustic features used for all experiments were 80-dimensional log-mel filter-banks computed on 25ms window with 10ms shift. Downsampling was performed through two convolutional layers with a factor of 4. Acoustic modeling units were the Chinese characters, which were 4233 and 6786 for AISHELL-1 and 10000-hour tasks respectively. For the training process, the Adam optimizer and the Noam \cite{vaswani2017attention} learning rate schedule were used, with warm-up steps of 25000. We also employed label smoothing and dropout regularization of 0.1 to prevent overfitting. Besides, the weight of the CTC loss was set to 0.3. 
All models were trained with 8 NVIDIA TESLA V100 GPUs, using the ESPNET toolkit \cite{watanabe2018espnet} with the PyTorch backend.

\subsection{AISHELL-1 Results}
For AHSHELL-1, speed-perturbation was applied to generate 3-fold versions of the data with factors of 0.9, 1.0 and 1.1. The model consisted of 12 encoder layers and 6 decoder layers. The feed-forward dimension was 2048 and the attention dimension was 256. Both the encoder and the decoder used multi-head of 4. During training, SpecAugment \cite{park2019specaugment} was applied on-the-fly to improve the robustness of the model. Models were trained for 50 epochs and we averaged the 10-best models based on the accuracy on development-set to generate the final models. Besides, two language models were trained using the transcriptions of the training set. A character-level two-layer LSTM-LM (650 memory cells for each layer) was used for beam search decoding, while a 4-gram word-level LM was employed in our WFST-based decoding process.

\subsubsection{Decoding performance}

\begin{figure}[t]
  \centering
  \includegraphics[width=0.9\linewidth]{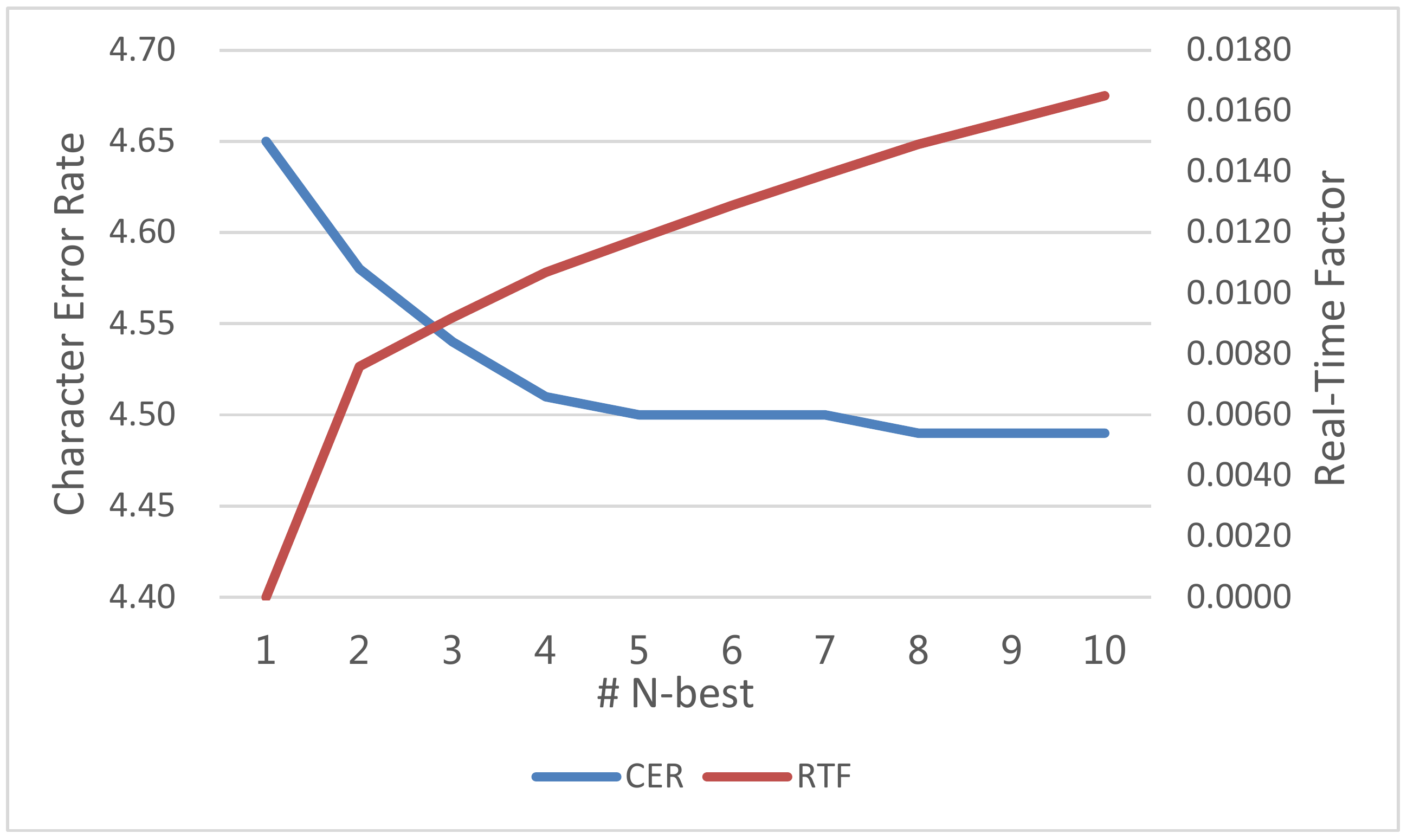}
  \caption{Curves of character error rate and real-time factor with the number of n-best.}
  \label{fig:nbest}
\end{figure}


We first examined the performance of our proposed WFST-based non-autoregressive decoding approach on a non-streaming model. Our first experiment evaluated the impact of the number of n-best hypotheses used for attention-decoder rescoring. Our observations are illustrated in Figure~\ref{fig:nbest}. As we can see, the CER drops significantly when n-best increases from 1 to 5 (1 means no rescoring applied), while the CER stays stable when nbest is greater than 5. However, the rescoring time was proportional to the number of n-best. In the following experiments we chose 5-best to balance the recognition accuracy and decoding efficiency.

\begin{table}[t]
  \caption{Decoding comparison}
  \label{tab:decoding_comparison}
  \centering
  \begin{tabular}{lll}
    \toprule
    \textbf{Decoding Strategy}   & \textbf{CER}   & \textbf{RTF}   \\
    \midrule
    Joint CTC-attention Beam Search                  & 4.7                & 0.74        \\
    \textbf{WFST-based Non-autoregressive}          & \textbf{4.5}       & \textbf{0.1}        \\
    \midrule
    CTC                          & 5.2                & -          \\
    CTC+LSTM-LM                  & 4.8                & -          \\
    \textbf{CTC+WFST}            & \textbf{4.65}      & \textbf{-}          \\
    \bottomrule
  \end{tabular}
\end{table}

Table~\ref{tab:decoding_comparison} shows the performance of the WFST-based non-autoregressive decoding approach on AISHELL-1 test-set. For comparison, the standard joint CTC-attention beam search method \cite{hori2017advances} was used as the baseline system. The character-level LSTM-LM was used during beam search progress in a shallow fusion manner. Hyper-parameters for beam search were: beam=10, ctc-weight=0.5, lm-weight=0.3. As can be seen in Table~\ref{tab:decoding_comparison}, the WFST-based non-autoregressive decoding approach achieves a CER of 4.5\%, with a relative improvement of 4.3\% compared to the baseline. Decoding efficiency of the two decoding strategy was also investigated in terms of RTF using a single thread with Intel(R) Xeon(R) E5-2630 v4 CPU, 2.20GHz. We can see the proposed method is 7.4x faster than the baseline. 

To evaluate the effectiveness of WFST-based word-level LM integration approach, we discarded the attention-decoder and only used the CTC branch for decoding. As for comparison, two baseline systems were built: one used CTC prefix beam search without any LMs, another used shallow fusion to integrate the character-level LSTM-LM during prefix beam searching process. We can see from Table~\ref{tab:decoding_comparison}, both LM integration approaches can improve the performance in CER. The shallow fusion approach with LSTM-LM improved the CER by 7.7\% in relative, while the word-level LM integrated with WFST brought a more improvement by 10.6\%.

Results in Table~\ref{tab:decoding_comparison} support the advantage of the WFST-based non-autoregressive decoding approach, both in speech recognition accuracy and efficiency. 

\subsubsection{Streaming processing}
Our streaming model was initialized with the non-streaming model to accelerate convergence. We employed multi-chunk strategy to process the input sequence as mentioned in section 3.2 and the chunk-size configuration was: \(\{N_{l}=(80,100,160),N_{c}=(32,48,64),N_{r}=(16,24,32)\}\). During training, \(N_{x}(x=l,c,r)\) randomly selected a number from the candidates, and a total of 27 combinations of \(\{N_{l},N_{c},N_{r}\}\) were generated.

\begin{table}[t]
  \caption{CERs with different Latency configuration on AISHELL-1 test-set}
  \label{tab:latency_comparison}
  \centering
  \begin{tabular}{llllll}
    \toprule
    \textbf{Model}   & \textbf{\(N_{l}\)}  & \textbf{\(N_{c}\)} & \textbf{\(N_{r}\)} & \textbf{latency(ms)} & \textbf{CER}   \\
    \midrule
    Non-streaming    & -   & -   & -  & -  & 4.5              \\
    \midrule
                    & \textbf{160} & \textbf{64} & \textbf{32} & \textbf{960} & \textbf{5.15}  \\
                    & 160   & 64   & 24  & 880  & 5.17       \\
                    & 160   & 64   & 16  & 800  & 5.21       \\
                    & 160   & 48   & 32  & 800  & 5.17       \\
    streaming       & 160   & 48   & 24  & 720  & 5.20       \\
                    & 160   & 48   & 16  & 640  & 5.25       \\    
                    & 160   & 32   & 32  & 640  & 5.22       \\
                    & 160   & 32   & 24  & 560  & 5.26       \\
                    & 160   & 32   & 16  & 480  & 5.33       \\
    \bottomrule
  \end{tabular}
\end{table}

In Table~\ref{tab:latency_comparison}, we investigated our WNARS system by varying the latency configurations. During inference, we fixed \(N_{l}=160\) as it has nothing to do with the model latency. As we can see, our best streaming model achieved a CER of 5.15\%, with a 960 ms latency and a 0.65\% absolute CER degradation compared with the non-streaming baseline. From the results in Table~\ref{tab:latency_comparison}, we can draw conclusions: 1) With the benefits of multi-chunk processing strategy, our WNARS system can achieve robust performance under different latency configurations. Only a 0.18\% absolute CER degradation were observed when latency reduced from 960ms to 480ms. 2) With the same latency, using longer \(N_{r}\) can bring better performance, as the model can take use of more future informantion.

Finally, in Table~\ref{tab:aishell_comparison}, we compared our WNARS system with several published streaming E2E solutions on AISHELL-1. For a fair comparison, the latency of the systems listed in Table 3 is around 600 ms. We can see our proposed method achieved the best result. To our best knowledge, this is the state-of-the-art performance for streaming ASR systems on this task.
\begin{table}[t]
  \caption{Comparison with published online-E2E systems on AISHELL-1}
  \label{tab:aishell_comparison}
  \centering
  \begin{tabular}{lll}
    \toprule
    \textbf{Model}   & \textbf{Latency(ms)}   & \textbf{CER}   \\
    \midrule
    SCAMA\cite{zhangstreaming}        & 600                & 7.39        \\
    MMA\cite{inaguma2020enhancing}    & 640                & 6.60        \\
    U2\cite{zhang2020unified}         & 640                & 5.42          \\
    \textbf{WNARS}                   & \textbf{640}       & \textbf{5.22}           \\
    \bottomrule
  \end{tabular}
\end{table}

\subsection{10,000-hour Results}
We verified the performance of the WNARS framework under a larger dataset, namely the 10,000-hour Mandarin dataset mentioned in \ref{ssec:data}. In this experiment, the attention dimension and the multi-head number of AED model was set to 384 and 6 respectively, while no speed-perturbation and SpecAugment were applied. As for comparison, we built a conventional ASR system with a TDNN-BLSTM lattice-free MMI (LFMMI) model \cite{qin2018automatic}. The model consisted of 9 TDNN layers (1536 hidden nodes for each layer with splice [-3, 0, 3]) and 4 BLSTM layers (1280 memory cells and 512 projections for each direction), with 4528 2-state context dependent phones as modeling units. After LFMMI training, sMBR \cite{kingsbury2009lattice} training was applied to obtain a strong baseline. Both systems employed an external 5-gram word-level LM during WFST-decoding process. 

As shown in Table~\ref{tab:10000-hour}, our WNARS system achieved a significant improvement in CER over the conventional ASR system, both on close-talk test-set and far-field test-set. Besides, the WNARS system has a much lower latency.
\begin{table}[t]
  \caption{Comparison with conventional ASR system on 10000-hour task}
  \label{tab:10000-hour}
  \centering
  \begin{tabular}{llll}
    \toprule
    \textbf{Model}   & \textbf{Latency(ms)}   & \textbf{close-talk}   & \textbf{far-field} \\
    \midrule
    LFMMI-sMBR        & 1900                & 8.2             &25.5    \\
    \textbf{WNARS}         & \textbf{960}       & \textbf{6.4}     & \textbf{20.4}         \\
    \bottomrule
  \end{tabular}
\end{table}

\section{Conclusions}
\label{sec:conclu}
In this work, we propose a novel WNARS framework to solve decoding efficiency, LM integration and streaming decoding problems of AED end-to-end model. With the help of WFST and multi-chunk strategy, our WNARS is robust under different inference latency. The proposed framework reaches a CER of 5.22\% with 640ms latency, which is the state-of-the-art streaming performance on AISHELL-1 task under comparable latency. Even with 50\% latency, WNARS still obtains over 20\% gain over a strong traditional industrial ASR system on a 10,000-hour Mandarin task, which certifies the effectiveness, practicability and generalization of our proposal. 

\section{Acknowledgements}
The authors gratefully acknowledge the financial support provided by the National Key Research and Development Program of China (2020AAA0108004).

\bibliographystyle{IEEEtran}

\bibliography{mybib}


\end{document}